\documentclass[doublecol,reprint,noshowpacs]{epl2}
\newcommand\dE{\text{d}E}
\renewcommand\d{\text{d}}

\newcommand{\ev}{{\vect{e}_{\Delta}}}

\usepackage{amsmath}
\usepackage{mathtools}
\usepackage{xcolor}
\usepackage{ulem}

\usepackage{amssymb}
\usepackage{placeins}

\title{Event-chain Monte Carlo algorithms for three- and 
many-particle interactions}

\author{J. Harland\inst{1} \and  M. Michel\inst{2,3} \and 
  T. A. Kampmann\inst{1} \and 
   J. Kierfeld\inst{1}}
\shortauthor{J. Harland \etal}

\institute{                    
  \inst{1} Physics Department, TU Dortmund University, 
      44221 Dortmund, Germany\\
  \inst{2} Orange Labs, 
   44 avenue de la R{\'e}publique, CS 50010, 
  92326 Ch{\^a}tillon CEDEX, France\\
  \inst{3} Laboratoire de Physique Statistique, 
   Ecole Normale Sup{\'e}rieure / PSL Research University, 
  UPMC, Universit{\'e} Paris Diderot, 
   CNRS - 24 rue Lhomond, 75005 Paris, France}

\abstract{ We generalize the rejection-free event-chain Monte Carlo
  algorithm from many-particle systems with pairwise interactions to
  systems with arbitrary three- or many-particle interactions.  We
  introduce generalized lifting probabilities between particles and
  obtain a general set of equations for lifting probabilities, the
  solution of which guarantees maximal global balance.  We validate
  the resulting three-particle event-chain Monte Carlo algorithms on
  three different systems by comparison with conventional local Monte
  Carlo simulations:
i) a test system of three particles with a three-particle interaction 
 that depends on the enclosed triangle area;
ii) a hard-needle system in two dimensions, where needle interactions
  constitute  three-particle interactions of the needle end points; 
iii) a semiflexible polymer chain with a bending energy, which constitutes 
a three-particle interaction of neighboring chain beads. 
The examples demonstrate that the generalization to many-particle
interactions broadens the applicability of 
event-chain algorithms considerably.
}

\begin{document}

\maketitle

\section{Introduction}

Monte Carlo (MC) simulations are (apart from Molecular Dynamics) the
main simulation technique for many-particle systems with a diverse
range of applications \cite{landau2015,frenkel2002}.  There has been
considerable progress on developing fast alternatives to the standard
local Markov-chain Monte Carlo (MCMC) technique, which is the
detailed-balance Metropolis algorithm.  Cluster MC algorithms are
non-local MCMC algorithms, where whole clusters of particles are moved
or updated within a single MC move.  For lattice spin systems, the
Swendsen-Wang \cite{Swendsen1987} and Wolff \cite{Wolff1989} have
provided the first cluster algorithms.  For off-lattice interacting
particle systems, the simplest of which are dense hard spheres,
different cluster algorithms have been proposed.  In ref.\
\citenum{Dress1995}, a cluster algorithm based on pivot moves has been
proposed \cite{Buhot1998,Santen2000,Liu2004}.  In ref.\
\citenum{krauth2009}, the event-chain (EC) algorithm has been
proposed, which provides a rejection-free algorithm where a chain of
hard particles is moved in each MC move
\cite{krauth2009,bernard2011,peters2012,michel2014,kampmann2014,isobe2015}.

The EC algorithm has been generalized \cite{michel2014, peters2012} to arbitrary
pairwise interactions \cite{kapfer2015} and continuous spin models
\cite{michel2015,nishikawa2015}.  In many applications, however,
three-particle interactions occur.  This happens, in particular, for
extended objects, such as rods or polymers, which can be described by
bead-spring models.  One prominent example are semiflexible polymer
chains with a  bending energy. Because the local bending angle
involves three neighboring beads in a discrete model, the bending
energy is a three-particle intra-polymer interaction in terms 
of bead positions.  Recently, the
EC algorithm has been applied to bead-spring models
of flexible polymer chains \cite{kampmann2015}. A
completely rejection-free algorithm for semiflexible polymers with
bending energy requires a rejection-free implementation of
three-particle interactions.

This is what we provide in the present paper. 
We will discuss how the EC approach can be generalized to 
arbitrary soft or hard three- and many-particle interactions. 
This generalization requires special lifting moves, because an EC
can transfer to {\it two}  (or more) 
possible interaction partners.  We provide 
a general solution of the set of lifting probabilities. 
We then validate and demonstrate the algorithm in three different 
applications. 
We start with a test problem involving only three particles with an 
interaction depending on the enclosed triangle area.
Then we proceed with hard needles in two dimensions. The steric interaction 
between needles can be formulated in terms of a three-particle hard-core 
interaction of their end points. 
Finally, we address the problem of a semiflexible polymer with the 
bending energy as three-particle interaction.

\section{Lifting probabilities}
\label{seq:3Part}

A MCMC algorithm produces a Markov chain, whose stationary
(unnormalized)
distribution $\pi$ is a Boltzmann distribution 
for a given system with energy $E$;
for a state $\alpha$, $\pi(\alpha)= \exp(-E(\alpha))$, where we
set $k_BT=1$, measuring energies in units of $k_BT$.
In order to retrieve the correct stationary
distribution, the algorithm has to fulfill the {\it global balance}
condition
\begin{equation}
\sum_\beta \phi(\beta\rightarrow \alpha)
=\sum_\gamma \phi(\alpha \rightarrow \gamma) =\pi (\alpha), 
\label{eq:GB}
\end{equation}
where $\phi(\alpha\!\rightarrow \!\beta)=\pi(\alpha)p(\alpha\!\rightarrow
\!\beta)$ is the probability flow from configuration $\alpha$ to $\beta$
and $p(\alpha\!\rightarrow\! \beta)$ the corresponding transition rate.
Instead of requiring {\it detailed balance} ($\phi(\alpha\!\rightarrow\!
\beta) = \phi(\beta\!\rightarrow\! \alpha)$ for {\it all} $\alpha,\beta$)
to fulfill (\ref{eq:GB}), EC algorithms satisfy {\it maximal global
  balance}, which means it is rejection-free
($\phi(\alpha\!\rightarrow\!\alpha)=0$) and flows between two
configurations are unidirectional ($\phi(\alpha\!\rightarrow\!\beta)> 0
\Rightarrow \phi(\beta\!\rightarrow\!\alpha)=0$).

For a system with $\mathcal{N}$-particle interactions, the total
energy $E= \sum_M E_M$ is the sum of all $\mathcal{N}$-body
interactions over all sets $M=\{i,j,...\}$ of $\mathcal{N}$ particles.
A move $\alpha\!\rightarrow\!\beta$ that involves displacements of one or
several particles generates corresponding energy changes $\Delta E_M$
in the interaction contributions, i.e., 
$\Delta E(\alpha\!\rightarrow\!\beta) = \sum_{M} \Delta E_{M}$.  
Detailed balance
is fulfilled by the standard Metropolis rule
$p^\text{Metr}(\alpha\!\rightarrow\!\beta) = \min\left(1,
    \pi(\beta)/\pi(\alpha)\right) =\min(1,\exp(-\Delta
E(\alpha\!\rightarrow\!\beta)))$ for acceptance of a move (offered with a
symmetric trial probability $p^\text{trial}(\alpha\!\rightarrow\!\beta)
=p^\text{trial}(\beta\!\rightarrow\!\alpha)$).  Factorizing the Boltzmann
weight along the sum of $\mathcal{N}$-particle interactions 
$\pi = \prod_M \pi_M$, we use a factorized Metropolis rule
\cite{michel2014},
\begin{align}
 p^\text{fact}(\alpha \rightarrow \beta)
=\prod_{M}  \min\left( 1, \frac{\pi_M(\beta)}{\pi_M(\alpha)} \right)
=e^{-\sum_{M}[\Delta E_M]^+}
\label{eq:FactMet}
\end{align}
where $[x]^+\equiv \max(0,x)$.

For infinitesimal moves with corresponding infinitesimal interaction
energy changes $\dE_M$, the probability of rejecting a move
$\alpha\!\rightarrow\!\beta$ simplifies further to
\begin{align}
1-p^\text{fact}(\alpha \rightarrow \beta)= 
\sum_{M} \frac{[-\d \pi_M]^+}{\pi(\alpha)}
=\sum_{M}[\dE_{M}]^+,
\label{eq:FactMet2}
\end{align}
which is simply the sum of all the positive contributions of the
$\mathcal{N}$-particle interactions \cite{michel2014}, called
  {\it factor}. 
A move can then be rejected by a single factor $M$ at a time.

\begin{figure}
\includegraphics[width=0.99\linewidth]{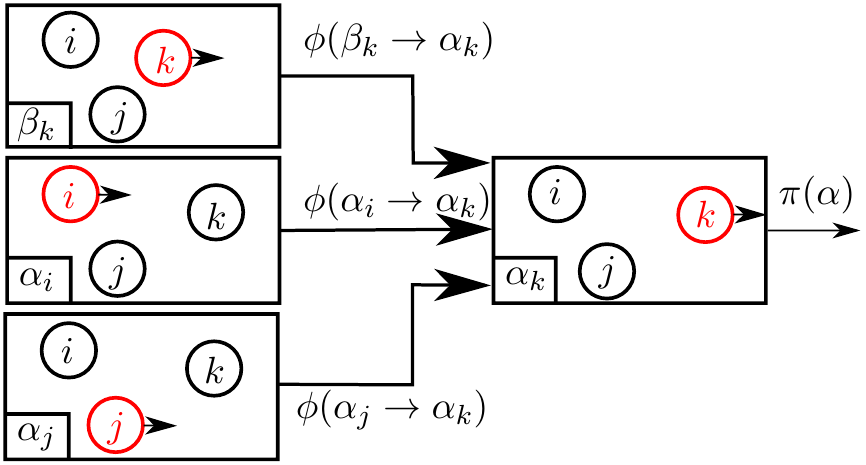}
\caption{
 Probability flow diagram for three particles. 
The flow into any configuration has
  to be equal to the outflow to satisfy (maximal) global balance.  Greek
  letters describe physical configurations and Latin letters denote lifting
  variables, i.e., moving particles.
}
\label{fig:flow_diagram}
\end{figure}

Drawing on the lifting framework \cite{diaconis2000}, maximal global
balance is enforced by extending the physical configurations $\alpha$
by a lifting variable $i$, which sets the particle for the next move,
to configurations  $\alpha_{i}$ (Greek letters
$\alpha,\beta,...$ describe the 
physical configuration, latin letters
$i,j,...$ the moving particle along the direction $\ev$).
We consider infinitesimal moves by vectors $\text{d}\vect{r}_i=\text{d}w \ev$,
first along a fixed unit vector direction $\ev$.
In the next move from the extended configuration $\alpha_{i}$
 particle $i$ is moved  by $\text{d}\vect{r}_i$
resulting in a new configuration $\beta_i$. 
 According
to the factorized Metropolis rule (\ref{eq:FactMet}), the physical
move $\alpha_{i}\!\rightarrow\!\beta_{i}$ is then either accepted
by all factors $M$ (the physical configuration $\alpha$ is
updated to $\beta$ and the next proposed move is again an update of
particle $i$ along $\ev$) or rejected by a single factor $M$. In
the latter case, a lifting move takes place, where the lifting
variable $i$ changes to another particle $k$ of the set $M$
resulting in a rejection-free algorithm.
We denote this lifting flow caused by the factor $M$ by 
$ \phi_{M}(\alpha_{i}\!\rightarrow\!\alpha_{k})$.

To obey global balance (\ref{eq:GB}),
lifting and physical flows into a configuration $\alpha_k$ must  
  add up to
its Boltzmann weight $\pi(\alpha)$
as  illustrated in fig.\ \ref{fig:flow_diagram},
\begin{align}
    \phi(\beta_{k}\rightarrow \alpha_{k})+  \sum_{M,k\in M} 
  \sum_{i\in M, i \neq k} \phi_{M}(\alpha_{i}\rightarrow
    \alpha_{k}) = \pi(\alpha)
\label{eq:flowK}
\end{align}
(for moves with fixed vectors $\text{d}w \ev$
only one configuration $\beta_k$ contributes in the sum in (\ref{eq:GB})).
Owing to detailed balance   
  of the factorized Metropolis filter 
 \eqref{eq:FactMet}, the physical flow $\phi(\beta_{k}\!\rightarrow\!
\alpha_{k})$ can be rewritten as
\begin{align}
\phi(\beta_{k}\rightarrow\alpha_{k}) 
  &= \pi(\beta)p^\text{fact}(\beta_{k}\rightarrow\alpha_{k}) 
  = \pi(\alpha)p^\text{fact}(\alpha_{k}\rightarrow\beta_{k})
   \nonumber\\
  &= \pi(\alpha)\left(1 - \sum_{M, k\in M} 
  [- \dE_{k,M}]^+ \right),
\label{eq:flowK2}
\end{align}
where  $+\dE_{k,M}$ is  the energy change of
  the set $M$ for a move of a particle $k\in M$ along $\ev$  
  for a   configurational change
  $\beta\!\rightarrow\! \alpha$ and 
 $-\dE_{k,M}$ the energy change for the reverse move of
  particle $k$. 
The lifting flow
  $\phi(\alpha_{i}\!\rightarrow\! \alpha_{k})$
 must compensate for the probability of rejection in \eqref{eq:flowK2}
to fulfill global balance  (\ref{eq:flowK}) without rejections.
We define the lifting probability 
$p^\text{lift}_M(\alpha_{i}\!\rightarrow\!\alpha_{k})$ from particle $i$ to $k$ 
within a factor $M$ by the decomposition 
\begin{align*}
\phi_{M}(\alpha_{i}\rightarrow \alpha_{k}) = \pi(\alpha)
   p^\text{lift}_{M}(\alpha_{i}\rightarrow\alpha_{k}).
\end{align*}

For infinitesimal displacements only a single  factor $M$ 
causes the
  rejection, i.e., 
 only the term from the rejection causing factor $M$ contributes 
  in the sum over lifting flows in the  global balance (\ref{eq:flowK}) 
(for a more detailed discussion see ref.\ \citenum{michel2015}). 
  Therefore,   it
  suffices to  consider  this specific rejection causing factor $M$ 
  for global balance in the following, so that
  global balance (\ref{eq:flowK}) is equivalent to 
\begin{align}
 \sum_{i\in M, i \neq k} p^\text{lift}_M(\alpha_{i}\rightarrow\alpha_{k}) =  
  [-  \dE_{k,M}]^+.
\label{eq:gbFact}
\end{align}

For pairwise interactions, see ref.\ \cite{michel2014}, each rejection
is caused by a single interacting particle and 
the factors $M$ in \eqref{eq:gbFact} are pairs. 
Due to
translational invariance, the resulting lifting probabilities  are
\begin{align}
p^\text{lift}_M(\alpha_{i}\rightarrow\alpha_{k}) = [-\dE_{k, i}]^+
  =[\dE_{i, k}]^+
\label{eq:pliftpair}
\end{align}
(with $\dE_{i,k} \equiv \dE_{i,M}$).
They give rise to lifting flows, which exactly 
 compensate for  rejections in \eqref{eq:flowK2}.
Maximal global balance is fulfilled:
There are no rejections on the extended configuration space and no
backwards moves on the physical space. Between two lifting moves, the particle
$i$ is moved by a finite displacement $\Delta w$, until a particle $k$
rejects the move and is moved in the same direction.

If the interaction is a many-body interaction, the lifting probability
(\ref{eq:pliftpair}) is not correct anymore and has to be adapted.
The translational invariance does not yield a symmetry between
  only two particles anymore but
between all particles in an interacting set $M$.
  Moving all particles $i\in M$  by the same
 $\text{d}\vect{r}_i=\text{d}w \ev$ leaves the energy $E_M$ invariant,
\begin{align}
\sum_{i \in M} \dE_{i, M} &= 0.
\label{eq:TransInv} 
\end{align}

In the following, we will discuss how to implement a maximal
global-balance scheme for many-body interactions, as illustrated in
fig.\ \ref{fig:flow_diagram}.
First, we decompose the overall lifting probability
  $p^{\text{lift}}_M(\alpha_i \to \alpha_k)$ into the trial probability
  to propose a lifting move from $i$ to any  particle of a factor $M$
  containing $i$ and $k$, $p^{\text{lift}}_M(\alpha_i \to \alpha_M)$, and
  into the conditional probability $\lambda_{ik}$ to actually lift
  from $i$ to $k$, so that $p^{\text{lift}}_M(\alpha_i \to \alpha_k) 
 = p^{\text{lift}}_M(\alpha_i \to \alpha_M) \lambda_{ik}$.
  In order to make the algorithm rejection-free, 
  the trial probability has to exactly compensate 
   the rejection probability $[\dE_{i, M}]^+$  
  for the rejection-causing 
  factor  $M$ containing $i$ from   \eqref{eq:FactMet2}, 
\begin{align*} 
p^{\text{lift}}_M(\alpha_{i} \to \alpha_M) = [\dE_{i, M}]^+,
\end{align*}
and the conditional probabilities have to 
be normalized: $\sum_{k\in M}\lambda_{ik} = 1$.

The global balance conditions \eqref{eq:gbFact} become
\begin{align}
   \sum_{i \in M} [\dE_{i, M}]^+ \lambda_{ik}=  
  [- \dE_{k,M}]^+.
  \label{eq:gbFact2}
\end{align}
Lifting from particle $i$ to $k$, i.e., 
$\lambda_{ik}> 0$  requires $\dE_{i, M}>0$
in order to trigger lifting by rejection 
and $\dE_{k, M}<0$ according to global balance
\eqref{eq:gbFact}.
This also  enforces
maximal global balance as only  lifting moves from $i$
to $k$ are proposed.
Let us consider a set $M$ of $\mathcal{N}$ interacting 
particles with $\mathcal{N}_-$ of them having 
$\dE_{k, M}<0$ (i.e., an update along $\ev$ leads to a decrease in energy)
 and $\mathcal{N}-\mathcal{N}_-$  having 
$\dE_{k, M}>0$  (i.e., 
 an update along $\ev$ leads to an increase in energy), 
for which we have to determine the set of 
 $(\mathcal{N}-\mathcal{N}_-)\mathcal{N}_-$ 
non-zero lifting probabilities $\lambda_{ik}$.  
The normalization $\sum_{k\in M\backslash \{i\}}\lambda_{ik} = 1$
gives $(\mathcal{N}-\mathcal{N}_-)$ conditions.
Global balance \eqref{eq:gbFact2} 
gives $\mathcal{N}_--1$ independent conditions (summing over $k$ 
leads to  $\sum_{i \in M} [\dE_{i, M}]^+ 
= \sum_{k \in M} [-\dE_{k, M}]^+$, which is always true 
 because of translational  invariance \eqref{eq:TransInv}). 
We thus have $\mathcal{N}-1$ independent conditions 
on $(\mathcal{N}-\mathcal{N}_-)\mathcal{N}_-$ 
non-zero $\lambda_{ik}$. We can conclude that 
for $\mathcal{N}=2,3$ these conditions are sufficient to obtain a unique 
set of $\lambda_{ik}$, whereas the choice of the probabilities
$\lambda_{ik}$ is not unique for $\mathcal{N}\ge 4$.

For $\mathcal{N}=2$ with two interacting particles $i$
and $k$, we simply have $\lambda_{ik}=1$, such that the EC algorithm
for pairwise interactions \cite{michel2014} is recovered.
For $\mathcal{N}=3$, i.e., three-particle interactions
global balance \eqref{eq:gbFact2} and normalization uniquely determine
 the  $\lambda_{ik}$ for the set 
$M=\{i,j,k\}$. If $\dE_{i,jk}>0$ (with $\dE_{i,jk} \equiv \dE_{i,M}$)
and with translational invariance
\eqref{eq:TransInv},   we have to 
distinguish three possible cases of signs of the 
energy changes $\dE_{j,ik}$ and $\dE_{k,ij}$:
\begin{align}
\begin{aligned}
\dE_{j,ik}&\!>\!0,~~\dE_{k,ij}\!<\!0: 
     && \lambda_{ij}=0,~\lambda_{ik}=1\\
\dE_{j,ik}&\!<\!0,~~\dE_{k,ij}\!>\!0:
     &&  \lambda_{ij}=1,~\lambda_{ik}=0\\
\dE_{j,ik}&\!<\!0,~~\dE_{k,ij}\!<\!0: 
   && \lambda_{ij}={[-\dE_{j,ik}]^+}/{[\dE_{i,jk}]^+}, \\
  & && \lambda_{ik}={[-\dE_{k,ij}]^+}/{[\dE_{i,jk}]^+}.
\end{aligned}
\label{eq:pij3particles}
\end{align}

For $\mathcal{N}\ge 4$, the choice of $\lambda_{ik}$ is not unique but 
there is a particularly simple choice, which we obtain with the 
additional conditions $\lambda_{ik} = \lambda_{jk}$  for all  $i,j \in M$.
Together with the global balance condition \eqref{eq:gbFact2},
we obtain
\begin{align}
  \lambda_{ik} = 
 \frac{[-\dE_{k, M }]^+}{\sum_{l \in M} [\dE_{l, M}]^+} = 
 \frac{[-\dE_{k, M }]^+}{\sum_{l \in M\backslash\{k\}} [\dE_{l, M}]^+}.
\label{eq:lambda}
\end{align}
Owing to  translational invariance, the normalization
 $\sum_{k\in M}\lambda_{ik} = 1$ holds, 
because the translational symmetry \eqref{eq:TransInv} 
leads to $\sum_{k \in M} [-\dE_{k, M}]^+ 
= \sum_{l \in M} [\dE_{l, M}]^+$.

Eq.\ \eqref{eq:lambda} is a  Glauber-like lifting rule as the  
lifting probability only depends on the energy 
change $[-\dE_{k, M}]^+$ of 
the final particle $k$. For the  case $\mathcal{N}=3$, the 
lifting rule 
  (\ref{eq:lambda}) is also equivalent to a Metropolis-like  representation
  $\lambda_{ik} = \min\left(1, [-\dE_{k,ij}]^+/
    [\dE_{i,jk}]^+\right)$.

The expression of the conditional probabilities is our main
  result: eq.\ (\ref{eq:lambda}) gives the rule to implement a maximal
  global-balance and rejection-free 
  scheme for $\mathcal{N}$-particle interactions
  provided the forces onto all $\mathcal{N}$ particles are known, as
  the infinitesimal energy changes $\dE$ correspond to the
  $\ev$-component of the forces on the particles.  
   For the  scheme  (\ref{eq:lambda}),
  the conditional lifting  probabilities $\lambda_{ik}$ depend on
  the MC-move direction $\ev$ and the forces onto the final particle $k$.

As for the EC algorithm for pairwise interactions, ergodicity on all
directions $\ev$ is achieved by setting a finite total displacement
$\ell$, \cite{peters2012,michel2014}. Once all the finite
displacements $\Delta w$ between successive lifting events sum up to
$\ell$, the lifting variable $i$ and the 
direction $\ev$ are resampled. 
This sequence is called an event chain, EC,
and $\ell$ the length of the EC. 

In practice, implementing infinitesimal moves leads to an infinite
number of physical moves per unit of time. An EC move starts with a
randomly chosen particle $i$ and a random direction
$\vect{e}_\Delta$. An event-driven approach is used to compute
directly the next lifting event. The maximal displacement length
$w_M$ for all many-particle interactions between any set $M$ 
of $\mathcal{N}$ particles containing $i$ are calculated by
solving
\begin{align}
\int_0^{w_M} [\dE_{i, M}(\tilde{w})]^+=- \ln u_M = \Delta E^*_{u_M},
\label{eq:DispLength}
\end{align} 
with $u_M$ being a random number uniformly distributed in $(0,1]$ and
drawn for each set $M$ such that the positive increment of energy
$\Delta E^*_{u_M}$ is drawn from an exponential distribution
\cite{peters2012}.
The particle is moved by the {\it smallest} $w=\min_{M} [w_M]$
selecting out one particular set of $\mathcal{N}$ interacting
particles for the lifting. Afterwards, the 
conditional lifting
probabilities $\lambda_{ij}$ for this set are calculated, and the EC
is lifted to the next moving particle accordingly. 
The computation of $\mathcal{N}-1$  lifting probabilities is not
 the performance-limiting step because 
the number of $\mathcal{N}$-particle tuples $M$, for which $w_M$ 
has to be calculated, is typically much larger.
Moving and lifting
are repeated until the EC length $\ell$ is reached.

\begin{figure}
\begin{center}
\includegraphics[width=0.85\linewidth]{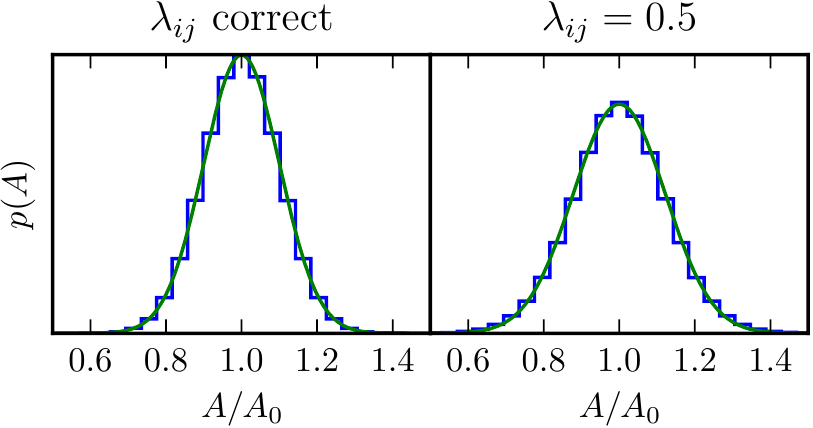}
\caption{
  Histogram of 
  measured areas with Gaussian fits. On the left we used the correct lifting
  ratios calculated by \eqref{eq:pij3particles}, while, on the right side, 
  we use the choice  $\lambda_{ij}=0.5$. 
   The systems were sampled with $K A_0^2=100$
such that the theoretical prediction \eqref{eq:TriangleGauss} is
   $\sigma_A=0.1\,A_0$ for the Gaussian distribution. 
We find $\sigma=(0.10068\pm0.0003)\,{A_0}$ for the correct solution
 but $\sigma=(0.1222\pm 0.0004)\, A_0$ for 
  $\lambda_{ij}=0.5$. 
}
\label{fig:TriangleDist}
\end{center}
\end{figure}


 For applications, the determination of the displacement  
$w_M$ is often one of the
 main technical difficulties. 
 As discussed in ref.\
  \cite{michel2014}, it is often advantageous to further decompose the
  interaction into several parts, e.g., $E_M =
  E_{a,M}+E_{b,M}+...$. Then, we treat each part as an independent factor
  in (\ref{eq:FactMet}) and determine for each part all maximal
  displacement lengths $w_{a,M}$, $w_{b,M}$,...\,.  The {\it smallest} $w
  = \min_{\{a,M\}}[w_{a,M}]$ gives the maximal displacement length and
  the
  conditional lifting probabilities $\lambda_{ij}$ are calculated
  for the set of particles $M$ and the part of the interaction $a$
  minimizing $w$.

  In the following sections, we validate our EC algorithm by applying
  it to three different systems with three-particle interactions.

\section{Triangle interaction}
\label{sec:Triangle}

 As a first validation we investigate 
a simple test system of three particles in two dimensions.
The three particles form a triangle of area $A$, and 
we define a genuine three-particle interaction by 
$E=\frac{K}{2}(A-A_0)^2$,
where $A_0$ is a preferred triangle area and $K$ is a coupling constant.  
The area can be written as 
$A=\frac{1}{2}|\vect{r}_{ij} \times \vect{r}_{ik}|$.

In order to calculate the maximal displacement length from eq.\ 
\eqref{eq:DispLength}, we need to analyze the 
 energy change $E(w)$ when moving particle $i$ along 
 $\Delta \vect{r}_i=w\vect{e}_\Delta$ for extrema. 
We find three zeros of $\dE(w)$,

\begin{equation*}
w_{01}=-\frac{|\vect{r}_{ij}(0)\times \vect{r}_{ik}(0)|}
   {(\ev \times \vect{r}_{jk})_z},~
 w_{02,03}=\frac{\pm 2A_0 } {(\ev \times \vect{r}_{jk})_z}+w_{01}.
\end{equation*}
At $w_{01}$ the area is zero, i.e.,
 all three particles are on a line,
 while $w_{02,03}$ are points where $A=A_0$. 
Solving \eqref{eq:DispLength} gives  the 
maximal displacement length
\begin{equation}
w= - w_{02,03} \pm  \left({\left(w_{02,03}\right)^2+
    {8\Delta E}/{K|\ev \times \vect{r}_{jk}|^2}}\right)^{1/2}
\end{equation}
where we use $w_{02}$ if $(\vect{r}_{ij}(0)\times \vect{r}_{ik}(0))_z<0$ 
and $w_{03}$ else and pick the smallest positive $w$. 
The lifting ratios are calculated by using \eqref{eq:pij3particles}.

The probability distribution of the area, $p(A)$, is given by the
Boltzmann distribution and should therefore be Gaussian,
\footnote{The number of triangles of area $A$, i.e., the 
accessible phase space of the 
three corner points of triangles
of area $A$  is independent of $A$.}
\begin{align}
p(A)=\frac{1}{Z}e^{- E(A)}  = \sqrt{\frac{K}{2\pi}} e^{-\frac{K}{2}(A-A_0)^2},
\label{eq:TriangleGauss}
\end{align}
with mean $\langle A \rangle =A_0$ and a variance $\sigma_A^2 =
K^{-1}$.
 In order to validate our algorithm we measure $p(A)$ and
compare with the theoretical prediction \eqref{eq:TriangleGauss}, see
fig.\ \ref{fig:TriangleDist}.  The correct choice of conditional
lifting probabilities $\lambda_{ij}$ agrees with the theoretical
prediction, whereas other choices such as $\lambda_{ij}=0.5$ give rise
to clear deviations.

\section{Hard needles in two dimensions}

\begin{figure}
\begin{center}
\includegraphics[width=0.8\linewidth]{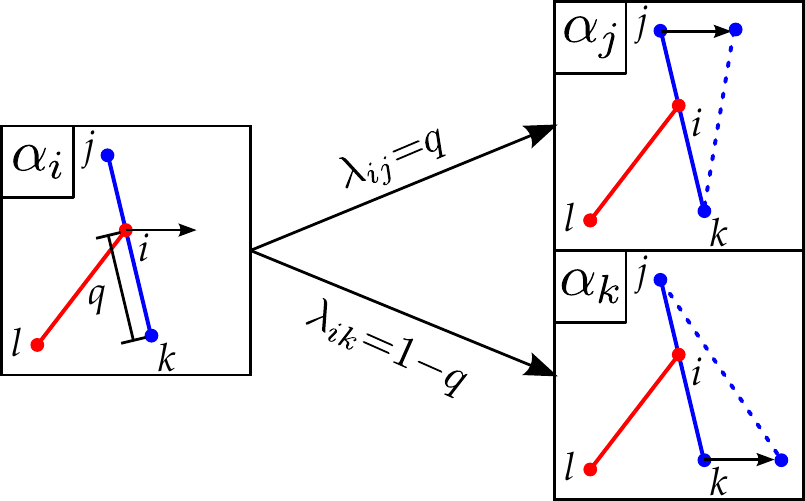}
\caption{
Lifting probabilities $\lambda_{ij}$ and $\lambda_{ik}$ when a moving end 
  point $i$  hits a fixed needle $jk$ (see eq.\eqref{eq:Needlelambda}).}
\label{fig:GB_Bonds}
\end{center}
\end{figure}

We now consider a system of hard extensible needles in two dimensions (2D), 
which are described in terms of the coordinates of their end points; 
the end point ensemble thus constitutes a many-particle system
that is treated with the EC algorithm.
Each needle $ij$ is extensible with a pair energy 
$E_{el} = \frac{1}{2} K(|\vect{r}_{ij}|-L)^2$, 
where $K$ is an elastic constant
and $L$ the needle rest length; we will focus on large $K$ to ensure
almost fixed length $L$.  The repulsive hard core  interaction of the lines
connecting the end points is modeled as three-particle interaction
between each end point $i$ and the line $jk$ connecting the two end
points $j$ and $k$ of any second needle, see fig.\ \ref{fig:GB_Bonds}. 
 In order to describe this
interaction we introduce the parallel component $r_{ki,||}=
\vect{r}_{ki}\cdot \vect{e}_{jk}$ of $\vect{r}_{ki}$ from $k$ to $i$
(using the unit vector $\vect{e}_{jk} \equiv
\vect{r}_{jk}/|\vect{r}_{jk}|$ parallel to the needle $jk$) and the
perpendicular distance $d_{i,jk} = |\vect{r}_{ki}-\vect{e}_{jk}
r_{ki,||}|$ of point $i$ to the needle $jk$. 

The Boltzmann weight of the hard needle interaction is zero if 
end point $i$ touches the needle $jk$ and one otherwise and 
can be written as
\begin{equation}
\pi_{i,jk} =
        1-  (1 - \Theta(d_{i,jk}))\Theta_L(r_{ki,||}),
\label{eq:pineedle}
\end{equation}
with  the Heaviside function $\Theta(x)$ ($=0$ for $x=0$ and $=1$ for $x>0$)
and another indicator function $\Theta_L(x)$ with $\Theta_L(x)=1$ for $0<x<L$ 
and $\Theta_L(x)=0$ otherwise.

Lifting occurs whenever an end point hits
another needle. This can happen either because the moving end point
hits a fixed needle or the moving needle  (belonging to the moving end
point) hits an end point of a fixed needle.  If an end point
$i$ hits the needle $jk$ the EC algorithm needs to decide whether to lift
to point $j$ or $k$, which gives the next end point to displace, see
fig.\ \ref{fig:GB_Bonds}.  For this decision we need to calculate the
conditional lifting probabilities $\lambda_{ij}$ and $\lambda_{ik}$,
given by \eqref{eq:pij3particles}. For the hard needle interaction, 
  we  use the derivative of the
  Boltzmann weights  $\dE_{i,jk}=-\d\pi_{i,jk}/\pi(\alpha)$ 
   (see eq.\ \eqref{eq:FactMet2}) with 
\begin{align*}
  &\frac{-\d\pi_{i,jk}}{\d w} =
     -(\nabla_{\vect{r}_i} d_{i,jk} \cdot \ev)  \delta(d_{i,jk}) 
                         \Theta_L(r_{ki,||})    \notag\\
 &  + (\nabla_{\vect{r}_i} r_{ki,||} \cdot\ev) 
     (\delta(r_{ki,||}) - \delta(L - r_{ki,||})) (1 - \Theta(d_{i,jk})).  
\end{align*}
The second term is non-zero only if the needle point $i$ exactly
  hits one of the ends of the needle $jk$ and can therefore be
  neglected for infinitely thin needles. 
 Using this in eq.\ \eqref{eq:pij3particles}
we find
\begin{align}
\lambda_{ik} =  
  \frac{  \left[ \nabla_{\vect{r}_k} d_{i,jk}\cdot \ev \right]^+ }
   {\left[-\nabla_{\vect{r}_i} d_{i,jk}\cdot \ev \right]^+ + 
    \left[-\nabla_{\vect{r}_j} d_{i,jk}\cdot \ev \right]^+  } 
\label{eq:Needlelambda}
\end{align}
with 
$\nabla_{\vect{r}_k} d_{i,jk} \cdot \ev = (\hat{\vect{r}}_{ik,\perp}\cdot \ev )
  |r_{ik,||}|/|r_{jk}|$ and 
$\nabla_{\vect{r}_i} d_{i,jk} \cdot \ev =
  -(\hat{\vect{r}}_{ik,\perp}\cdot\ev )$.

When the moving end point  $i$ hits the needle $jk$, this simplifies to 
 simple length ratios $\lambda_{ij} = |r_{ik,||}|/|r_{jk}|\equiv q$
and  $\lambda_{ik} = |r_{ij}|/|r_{jk}|= 1-q$, where $|r_{jk}|\approx L$. 
When, vice versa,  point $k$ is moving and 
 an end point $i$ is hit by the moving needle $jk$, it gives
   $\lambda_{ki} = 1$ and $\lambda_{kj} =0$, i.e., the 
EC then transfers to $k$ with certainty.
The elastic  pair energy $E_{el}$ is treated independently as an 
additional    simple pairwise
interaction  using  eq.(\ref{eq:pliftpair}).

\begin{figure}
\begin{center}
\includegraphics[width=.99\linewidth]{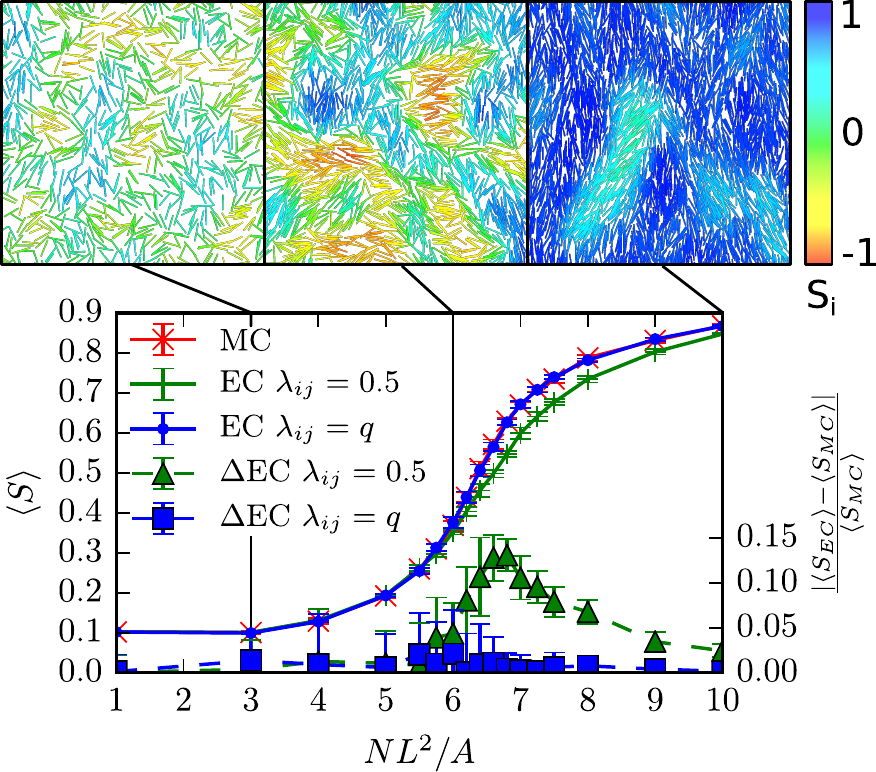}
\caption{ 
  Nematic order parameter $\langle S \rangle$  (left axis) and relative 
  deviation of EC results to local MC (right axis) 
  as a function of the 
   number of needles $N$ per area $A$ for a 2D system
  of needles with  $K=100/L^2$ 
  with periodic boundaries ($A=10L\times10L$) (left axis). 
  The EC algorithm  with $\lambda_{ij} = q = 1-\lambda_{ik}$, see eq.\ 
  (\ref{eq:Needlelambda}), agrees numerically with local  MC,
    whereas a naive choice
  $\lambda_{ij}=\lambda_{ik}=0.5$ 
  significantly deviates. Upper row: Simulation snapshots color-coded 
   for the local order parameter 
    $S_i=2 \cos^2\theta_i-1$ around needle $i$.
}
\label{fig:LC_OP}
\end{center}
\end{figure}

In order to validate our algorithm
we measure  the nematic order parameter of the 2D
hard needle system
$S=2\langle\cos^2\theta\rangle-1$,
with $\theta$ being the angle between the needle orientation and the 
director
as a function of needle density.
In fig.\ \ref{fig:LC_OP}, we compare 
local MC  simulations with rejections  and
two versions of the  EC algorithm.
In one version we naively take $\lambda_{ij}=\lambda_{ik}=1/2$,
the other version is the proper algorithm using (\ref{eq:Needlelambda})

Measuring the autocorrelation time for the order parameter $\langle S\rangle$
we find speed-up factors of $3-4$ in CPU time 
for the EC algorithm in comparison to the
local MC algorithm (measured at $NL^2/A = 6$). 
This algorithm can also be used in
  polymer simulation where the polymers are modeled as chains of hard needles
  as alternative to existing polymer EC algorithms \cite{kampmann2015}.

\section{Semiflexible polymer}

\begin{figure}[t]
\begin{center}
\includegraphics[width=0.85\linewidth]{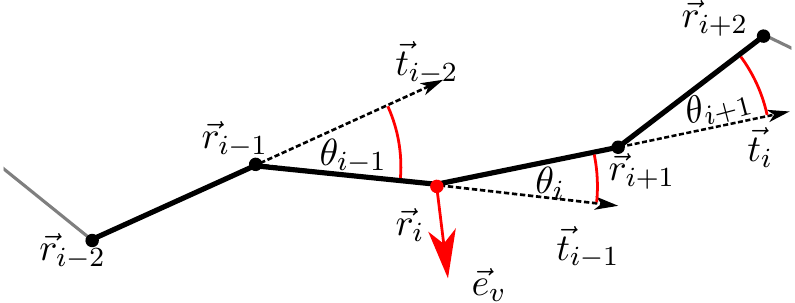}
\caption{
  Moving bead $i$ in a semiflexibe polymer changes three angles.
}
\label{fig:SemiFlexPoly}
\end{center}
\end{figure}

As a third application
we simulate a free 
semiflexible harmonic chain with bending rigidity $\kappa$
  composed of $N$ beads and elastic bonds of 
mean length $b=1$ \cite{kierfeld2004}, i.e. mean contour length $L=Nb$,
 in three dimensions.\footnote{The harmonic bond stretching 
  energy is handled as in  \cite{kampmann2015} as additional 
  pair interaction in the EC algorithm.}  
The  bending energy is 
\begin{equation}
E_{b}=\kappa \sum_{i=1}^{N-2} 
    \left(1-\frac{\vect{t}_i\cdot \vect{t}_{i+1}}
    {|\vect{t}_i||\vect{t}_{i+1}|}\right)
   =\kappa \sum_{i=1}^{N-2}\big(1-\cos(\theta_{i+1})\big)
\label{eq:HWLC}
\end{equation}
with tangential vectors $\vect{t}_i=\vect{r}_{i+1}-\vect{r}_{i}$ and
tangential angles $\theta_i$  at bead $i$. Moving
bead $i$ changes three terms in \eqref{eq:HWLC}  since $\vect{t}_{i-1}$,
$\vect{t}_i$ and $\vect{t}_{i+1}$ are functions of $\vect{r}_i$ and three
angles change (see fig.\ \ref{fig:SemiFlexPoly}). 
Each angle is a function of three particle positions,
making the bending energy  a three-particle interaction. 
MC algorithms working on bead positions rather 
than  angles are important for the simulation of 
many-polymer systems  or polymers in external potentials, 
where interactions are position-dependent.

We first determine the maximal displacement length 
for a displacement $\Delta \vect{r}_i=w\vect{e}_\Delta$ of bead $i$.
In the following, $x$ is the remaining total displacement length of the
EC, i.e. $w\le x$. 

\begin{figure}[t]
\begin{center}
\includegraphics[width=.99\linewidth]{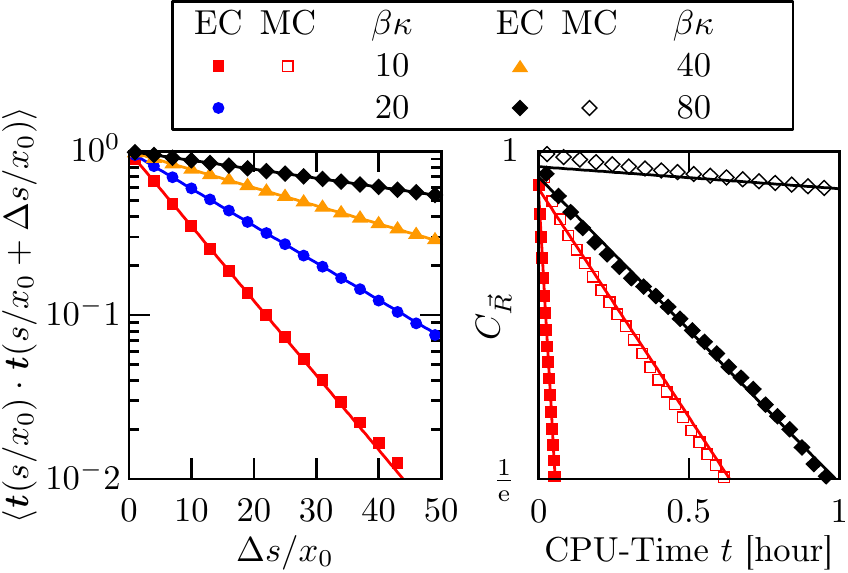}
\caption{
   Left: Tangent correlation function and expected exponential decay (dashed)
  for different values of $\kappa$ and $N=240$ beads.  
Right: Autocorrelation $C_{\vect{R}}(\Delta t)$ 
  of the end-to-end-vector as a function of CPU time $t$ 
(on a $3.7$GHz Intel Xeon CPU)
 for local MC and the new EC algorithm. 
Lines are 
exponential fits $\sim e^{-t/\tau}$.
}
\label{fig:TangAutoKorr}
\end{center}
\end{figure}

For  an outer angle $\theta_{i-1}$ the  energy change  is
\begin{align}
\dE_{i,\theta_{i-1}}(w)&=
   -\frac{\kappa}{|\vect{t}_{i-2}||\vect{t}_{i-1}+w
     \vect{e}_\Delta|^3}
   (a + w b) \text{d}w,
\nonumber
\\
 a&\equiv    \big( \vect{t}_{i-1}^2 (\vect{t}_{i-2}\cdot \vect{e}_\Delta) -
    (\vect{t}_{i-1}\cdot \vect{t}_{i-2})(\vect{t}_{i-1}\cdot\vect{e}_\Delta)
    \big) 
\nonumber\\
 b   &\equiv \big( (\vect{t}_{i-1}\cdot\vect{e}_\Delta)
     (\vect{t}_{i-2}\cdot \vect{e}_\Delta) -
(\vect{t}_{i-1}\cdot\vect{t}_{i-2})
    \big),
\nonumber
\end{align}
which gives  an extremum   of the energy $E_{i,\theta_{i-1}}(w)$ at 
 $w_0= - a/b$. 
 For the  calculation of the maximal displacement length we calculate
 $w_0$ and  the sign of 
$\dE_{i,\theta_{i-1}}(w=0)$. 
If $\dE_{i,\theta_{i-1}}(w=0)<0$,  $E_{i,\theta_{i-1}}(w)$ is decreasing.
If $w_0<0$ the bead can move until $w=x$  is reached
(and the EC terminates). 
If $w_0>0$
 there is an energy minimum at $w_0$, to which 
 the bead
can move  at no energy cost. We calculate  the 
 energy $E_\text{\rm min}$  and 
  allow an energy increase of $\Delta E$  to 
$E_\text{\rm new}=E_\text{\rm min}+\Delta E$.
The corresponding 
 displacement length $w_+$  is calculated by
solving
\begin{align*}
\cos{\theta_+} \equiv 1-\frac{E_\text{\rm new}}{\kappa} 
  =\frac{\vect{t}_{i-2}\cdot 
 (\vect{t}_{i-1}+w_+\vect{e}_\Delta)}
    {|\vect{t}_{i-2}||\vect{t}_{i-1}+w_+\vect{e}_\Delta|},
\end{align*}
which gives
\begin{equation}
w_+=-{B}/{A}\pm\sqrt{\left({B}/{A}\right)^2-{C}/{A}},
\label{eq:SFw+}
\end{equation}
\begin{align*}
A & \equiv (\vect{t}_{i-2})^2\cos^2(\theta_+)-(\vect{t}_{i-2}\cdot
\vect{e}_v)^2,  \nonumber\\
B & \equiv (\vect{t}_{i-1}\cdot
\vect{e}_v)(\vect{t}_{i-2})^2\cos^2(\theta_+)
   -(\vect{t}_{i-2}\cdot\vect{t}_{i-1})(\vect{t}_{i-2}\cdot\vect{e}_v),
    \nonumber\\
C  & \equiv (\vect{t}_{i-1})^2(\vect{t}_{i-2})^2\cos^2(\theta_+)
-(\vect{t}_{i-2}\cdot\vect{t}_{i-1})^2,
  \nonumber
\end{align*}
where we take the smallest positive solution in \eqref{eq:SFw+},
and the maximal  displacement length  is $w=w_0+w_+$.
If ${\text{d}E_{i,\theta_{i-1}}(w=0)>0}$, $E_{i,\theta_{i-1}}(w)$ is
increasing and there is an energy minimum at $w_0$.
Then we calculate  $w_+$ immediately using \eqref{eq:SFw+} with 
$E_\text{\rm new}=E_{i,\theta_{i-1}}(0)+\Delta E$. 
 If  $w_+$ is smaller than the energy-maximizing $w_0$ or
$w_0<0$, $w_+$ is the maximal displacement length.
 Otherwise, the move over the energy maximum at $w_0$ can be performed
so the bead can move until $w=x$  is reached
(and the EC terminates). 
 Analogous
calculations apply to  $\theta_{i+1}$.

The calculation for the center angle $\theta_i$  is more complicated
because $\text{d}E_{i,\theta_i}(w)$ has one or three zeros. Since the energy 
$E_{i,\theta_i}(w)$ gets maximized for $w\rightarrow \pm \infty$, 
 $E_{i,\theta_i}(w)$ either has two minima and one maximum in between 
or only a single  minimum. 
The algorithm now works as follows: First the zeros $z_{i}$ with
$\text{d} E_{i,\theta_i}(z_i) =0$ are calculated 
 Since only zeros in the moving direction are important there can be up to
 three extrema on the way.  These four cases are treated as follows:\\ 
{\it 0   zeros}: $\text{d}E_{i,\theta_i}>0$ so moving costs energy.
 The energy cost
 for moving the complete remaining $x$ can be calculated. If the cost is
 smaller than $\Delta E$ this move is performed.  Otherwise, $w$ is calculated
 numerically by solving $E_{i,\theta_i}(w) = E_{i,\theta_i}(0)+\Delta  E$.\\ 
{\it 1 zero}: $z_{1}$ is a minimum. Set $x = x-z_{1}$ and go to case
 no zeros as explained above.\\ 
{\it 2 zeros}: The first zero $z_{1}$ is a maximum,
 the second one $z_{2}$ a minimum. If the maximum can be reached with the
 energy $\Delta E$, the available energy $\Delta E$ is reduced by the energy
 cost $\Delta E = \Delta E - (E_{i,\theta_i}(z_{2})-E_{i,\theta_i}(z_{1}))$.
 Set $x = x-z_{1}$ and continue at case 1 zero as above with $z_{1}=
 z_{2}-z_{1}$ If the maximum at $z_{1}$ cannot be reached, $w$ is calculated
 numerically on the interval $[0,z_{1}]$.\\ 
{\it 3 zeros}: Then, $z_{1}$ is
 a minimum, which means the particle can move to $z_{1}$.  Set $x = x-z_{1}$
 and continue at case 2 zeros as above with $z_{1}= z_{2}-z_{1}$ and
 $z_{2}= z_{3}-z_{1}$.

We validate the correctness of this algorithm 
by measuring the tangent correlation function
$\langle \vect{t}(s)\cdot\vect{t}(s+\Delta s)\rangle$, 
which is analytically known to decay as  $e^{-\Delta s/L_p}$ 
(for $\Delta s \ll N$)
with the persistence length 
$L_p=-1/\ln(1/\tanh \kappa -1/\kappa)$ 
\cite{kleinert2006}. As seen in fig.\ \ref{fig:TangAutoKorr} the measured
values match the analytical curve.

To measure the efficiency of the algorithm the autocorrelation of the
end-to-end-vector $\vect{R}=\vect{r}_N-\vect{r}_0$ is calculated as
\begin{align}
C_{\vect{R}}(\Delta t)=
\frac{\langle\vect{R}(t)\cdot \vect{R}(t+\Delta t)\rangle-
   \langle\vect{R}\rangle^2}
{\langle\vect{R}^2\rangle-\langle\vect{R}\rangle^2} 
  \underset{\Delta t \gg 1}{\sim}  e^{-\Delta t/\tau},
\end{align}
where the time is measured in real time to allow a comparison of the
  EC algorithm with the local MC method  The results are
shown in fig.\ \ref{fig:TangAutoKorr}.   For $\beta \kappa = 10$ we get
  $\tau_{\rm MC} \approx 0.72{\rm h}$ and $\tau_{\rm EC} = 0.06 {\rm h}$,
  which means the EC algorithm is approx.\ $11.3$ times faster at
  equilibrating a semiflexible polymer than the standard local MC
  method.  For $\beta \kappa = 80$ the EC performs even better: we
  get $\tau_{\rm MC} \approx 15{\rm h}$ and $\tau_{\rm EC} = 1.1 {\rm h}$,
  which gives a speed-up of approx.\ $14$.

\section{Discussion and Conclusion}

We generalized the EC algorithm to three-particle and 
many-particle interactions thus broadening the range of applicability 
of rejection-free EC algorithms considerably. 
For  $\mathcal{N}$-particle interactions, there are $\mathcal{N}-1$ 
interacting particles to which the EC can lift to 
avoid rejections. We calculate a set of $\mathcal{N}-1$ 
conditional lifiting probabilities 
$\lambda_{ij}$  which assure maximal global balance. 

We applied the generalized EC algorithm successfully to 
three different systems -- a small system with 
three  particles with a triangle-area-dependent  interaction, 
hard needles in 2Ds, and a single semiflexible polymer chain 
with bending energy -- and demonstrate in all three cases the correctness
of the algorithm. For hard needles or the semiflexible polymer
we obtain considerable performance 
gains. 

In the future, the EC algorithm can be used for efficient 
 large scale simulations of the 2D hard needle system to 
answer the questions as to whether nematic long-range order exists in large 
systems (the increase of $S$ to a large value in 
 fig.\ \ref{fig:LC_OP} can be an artefact of finite size
effects)
 and  whether the system has a 
 Kosterlitz-Thouless transition around $\rho \sim 7/L^2$ 
by disclination unbinding, as suggested by local MC results 
\cite{frenkel1985,vink2009}.
Future work should also evaluate systematically to what extent
the EC algorithm can suppress critical slowing down at the transition 
from isotropic to (quasi-)nematic. 

With respect to polymer simulations, the algorithm will allow the 
entirely
rejection-free simulation of large systems containing many interacting
semiflexible polymers \cite{kampmann2015,kampmann2014}.  In our previous
  work\cite{kampmann2015} hybrid EC algorithms were slower than algorithms
  where all interactions are handled by the EC scheme, so that we
  expect that the simulation of  many-polymer systems with bending energies 
  will also benefit from
  the $\mathcal{N}$-particle EC algorithm.



\end{document}